\documentclass{article}
\PassOptionsToPackage{table, dvipsnames}{xcolor}
\usepackage{arxiv}
\usepackage[utf8]{inputenc} 
\usepackage[T1]{fontenc}    
\usepackage{url}            
\usepackage{nicefrac}       
\usepackage{soul}
\usepackage{fontawesome}
\usepackage[many]{tcolorbox}
\usepackage[table]{xcolor}
\usepackage{booktabs} 
\usepackage{lipsum}
\usepackage{xspace}
\usepackage{enumitem}
\usepackage{natbib}
\usepackage{xcolor}
\definecolor{darkblue}{RGB}{44,62,80}
\definecolor{CalGoldHex}{RGB}{253, 181, 21} 
\definecolor{gold}{RGB}{149, 113, 30} 
\usepackage{tikz}
\usetikzlibrary{arrows.meta,positioning,fit,calc,shapes.multipart}

\usepackage[breaklinks=true,colorlinks,bookmarks=false,citecolor=gold,linkcolor=darkblue]{hyperref}

\usepackage{enumitem}
\usepackage{doi}

\usepackage{amsmath}
\usepackage{amssymb}
\usepackage{mathtools}
\usepackage{amsthm}

\usepackage[capitalize,noabbrev]{cleveref}

\usepackage{fourier}

\usepackage[textsize=tiny]{todonotes}

\usepackage{tabularx}

\title{Prioritization First, Principles Second: An Adaptive Interpretation of Helpful, Honest, and Harmless Principles}

\newcommand\blfootnote[1]{
  \begingroup
\renewcommand\thefootnote{}\footnote{#1}%
  \addtocounter{footnote}{-1}%
  \endgroup
}

\newcommand*{\affmark}[1][*]{\textsuperscript{\textnormal{#1}}}

\author{
\textbf{Yue Huang}\affmark[1], \textbf{Chujie Gao}\affmark[1], \textbf{Yujun Zhou}\affmark[1], \textbf{Kehan Guo}\affmark[1], \textbf{Xiangqi Wang}\affmark[1], \textbf{Or Cohen-Sasson}\affmark[2]\\ \textbf{Max Lamparth}\affmark[3] \textbf{Xiangliang Zhang}\affmark[1]\\
~\\
\affmark[1]Department of Computer Science and Engineering, University of Notre Dame~~~
\\\affmark[2]School of Law, University of Miami, \affmark[3]Center for AI Safety, Stanford University
}

\renewcommand{\shorttitle}{Prioritization First, Principles Second: An Adaptive Interpretation of Helpful, Honest, and Harmless Principles}

\usepackage{pythonhighlight}
\usepackage{booktabs}
\usepackage{graphicx}
\usepackage{booktabs}
\usepackage{caption}
\usepackage{subcaption}
\usepackage{fourier}
\usepackage{multirow}
\usepackage{comment}
\usepackage{wrapfig}

\usepackage{enumitem}
\usepackage{array}

\newcolumntype{C}[1]{>{\centering\arraybackslash}m{#1}}
\newcolumntype{L}[1]{>{\raggedright\arraybackslash}m{#1}}

\usepackage{color}
\definecolor{darkblue}{RGB}{44,62,80}
\definecolor{CalGoldHex}{RGB}{253, 181, 21} 
\definecolor{gold}{RGB}{149, 113, 30} 
\usepackage{tcolorbox} 

\tcbset{
  definitionbox/.style={
    colback=green!5!white,
    colframe=black!50,
    boxrule=0.5pt,
    sharp corners,
    left=1.5mm,
    right=1.5mm,
    top=1mm,
    bottom=1mm,
  }
}
\usepackage[breaklinks=true,colorlinks,bookmarks=false,citecolor=gold,linkcolor=darkblue]{hyperref}

\definecolor{deepred}{rgb}{0.631,0.102,0.102}
\definecolor{amethyst}{rgb}{0.6, 0.4, 0.8}
\definecolor{darkgreen}{rgb}{0.3,0.7,0.3}
\definecolor{salmon}{RGB}{241, 150, 141}

\begin{document}
\maketitle
\blfootnote{Correspondence to: Yue Huang~(\url{yhuang37@nd.edu}).}

\begin{abstract}
The Helpful, Honest, and Harmless (HHH) principle is a foundational framework for aligning AI systems with human values. However, existing interpretations of the HHH principle often overlook contextual variability and conflicting requirements across applications. In this paper, we argue for \textbf{an adaptive interpretation of the HHH principle} and propose a reference framework for its adaptation to diverse scenarios. 
We first examine the principle's foundational significance and identify ambiguities and conflicts through case studies of its dimensions. To address these challenges,  we introduce the concept of priority order, which provides a structured approach for balancing trade-offs among helpfulness, honesty, and harmlessness.  Further, we explore the interrelationships between these dimensions, demonstrating how harmlessness and helpfulness can be jointly enhanced and analyzing their interdependencies in high-risk evaluations.
Building on these insights, we propose a reference framework that integrates context definition, value prioritization, risk assessment, and benchmarking standards to guide the adaptive application of the HHH principle. This work offers practical insights for improving AI alignment, ensuring that HHH principles remain both ethically grounded and operationally effective in real-world AI deployment.
\end{abstract}

\section{Introduction}

The development of AI assistants has progressed rapidly, evolving from simple rule-based systems \citep{10.1145/365153.365168} to advanced models, such as Large Language Models (LLMs) \citep{zhao2023survey}. The wide application of these models highlights not only the expanding capabilities of AI but also the growing need to ensure these systems align with human preferences and values. To address this challenge, the HHH principle--standing for Helpful, Honest, and Harmless--was proposed by \citet{askell2021general} as a guiding framework for designing and evaluating AI systems. The principle aims and \textit{has been widely utilized} to align AI behavior with human-centered values, providing critical benchmarks for tasks such as training data selection, strategy design, and deployment guidance \citep{touvron2023llama, li2024llava, bai2023qwen}. Its adoption has been instrumental in advancing AI alignment, particularly in ensuring that powerful models prioritize user benefit, truthfulness, and safety.

However, current studies indicate the existence of ambiguities and conflicts among the three HHH dimensions. For example, the same prompt input might be classified as either harmless or harmful depending on the context or criteria used in different research. Similarly, the definition of specific dimensions often varies across studies, influenced by the diverse contexts in which they are applied. Furthermore, key questions such as ``which dimension should be prioritized'' and ``how different dimensions interrelate'' remain unresolved, underscoring the need for a clearer and more adaptable interpretation of the HHH principle.

In this position paper, \textbf{we argue that the HHH principle is not static or rigid but requires an adaptive interpretation to remain effective across diverse scenarios}. 
To support this argument, first, we review the initial definition and fundamental importance of the HHH principle, examining its positive impact on AI alignment efforts (\textbf{\S\ref{sec:background}}). Next, we identify ambiguities and conflicts within current interpretations of the principle through case studies of different dimensions (\textbf{\S\ref{sec:ambiguity}}), highlighting the need for consensus or innovative solutions. To address these conflicts, we introduce the concept of priority order, which provides a structured approach to balancing competing requirements among dimensions (\textbf{\S\ref{sec:priority_order}}). Furthermore, we investigate the relationships between dimensions, analyzing how harmlessness and helpfulness can be simultaneously improved and examining their interdependencies in high-risk evaluations (\textbf{\S\ref{sec:relation}}). Building on these insights, we propose a reference framework that guides the adaptive application of the HHH principle to specific scenarios (\textbf{\S\ref{sec:framework}}). The framework outlines critical considerations and offers a comprehensive methodology for interpreting and operationalizing the HHH principle in varied contexts.

Through this position paper, we aim to deepen the understanding of the HHH principle and foster its effective utilization as a tool for advancing AI development and alignment. We hope this work inspires further research and discussion on aligning AI systems with human values in a nuanced and adaptable manner.

\vspace{-7pt}
\section{Fundamental Value of the HHH Principle}
\label{sec:background}

The HHH principle is a set of guiding principles for developing and aligning AI models, particularly large language models, with human values. \citet{askell2021general} define it from its aim: \textit{We will define an AI as "aligned" if it is, in three words, helpful, honest, and harmless or 'HHH'. Our alignment efforts aim to measure and address this general problem with large language models.} For the sake of simplicity, we summarize the HHH principle based on their original definitions: 

\vspace{5pt}
\textit{1) \textbf{Helpful}:  AI models should assist users by providing useful, accurate, and contextually relevant information or services. They must be designed to meet user needs, enhance productivity, and effectively solve problems. \\
2) \textbf{Honest}:  AI models should ensure transparency and truthfulness in their responses, providing factual information while openly acknowledging their limitations. They must refrain from generating falsehoods or misleading content. \\
3) \textbf{Harmless}:  AI models should avoid causing harm by preventing the generation of biased, offensive, or unethical content. They should prioritize safety and respect in their interactions, ensuring that they do not produce harmful or inappropriate outputs.}
\vspace{5pt}

\textbf{The HHH principle plays a critical role in guiding   AI system development to align with human values and preferences.} This principle has been used for 
establishing clear, user-centric benchmarks \citep{zheng2023judging}, and designing training methods like Reinforcement Learning from Human Feedback (RLHF) \citep{ouyang2022training} for advancing AI system performance. \textbf{Helpfulness} is directly related to the functional effectiveness of AI models, addressing what \textit{human users really need}, thus enhancing human productivity. \textbf{Honesty} ensures that AI systems provide truthful (\emph{e.g.}, avoid hallucination \citep{huang2023survey, zhang2023siren}), and transparent (\emph{e.g.}, express uncertainty \citep{xiong2024can}) information, which is crucial for fostering trust between humans and machines. \textbf{Harmlessness}, in turn, guarantees that AI outputs are safe and ethical, preventing the generation of harmful, biased, or misleading content \citep{huang2024position, decodingtrust, liu2023trustworthy}. The HHH principle determines the three most important sub-directions aligned with human preferences and values, which have been widely used for guiding AI model design as shown in \autoref{app:model_example}.

\textbf{The HHH principle uniquely emphasizes both trustworthiness and utility \footnote{Utility here refers to the model effectiveness  in natural language processing tasks, including  logical reasoning, content summarization, text generation, and so on \citep{huang2024position}.}.} 
AI models, such as LLMs, interact with users by providing information, generating content, and assisting in decision-making, making trustworthiness and utility essential for their effectiveness and reliability. While utility is prioritized in model development through training and tuning loss minimization, an automated and human-effortless process, there is a risk of compromising ethical integrity or introducing biases. Thus, typical alignment principles emphasize additional efforts to align AI models with human values, as shown in \autoref{app:model_example}.  Research indicates that the resulting models often choose safe responses over helpful ones \citep{touvron2023llama1}. However, the HHH principle offers a unique balance by explicitly emphasizing both dimensions. The dimension of \textbf{honesty} requires AI models generate accurate and truthful outputs. 

\textbf{The HHH principle facilitates the transition of AI from a passive tool to an active participant.} The societal impact of AI models extends \textit{beyond tangible benefits}, influencing more subtle and indirect aspects of human life. AI-generated content has the potential to shape human perspectives (\emph{e.g.}, create ideas in scientific research \citep{si2024can} and simulate social behaviors \cite{huang2024social}), cultural norms, and even values \citep{ramezani-xu-2023-knowledge, agarwal-etal-2024-ethical}. Repeated exposure to biased or subtly manipulative outputs could inadvertently influence public opinion or entrench societal biases \citep{zeng2024ai}. This transition from AI being a \textit{passive tool} to an \textit{active participant} raises concerns about its potential role as a manipulative force in shaping discourse and decision-making  \citep{pmlr-v235-simmons-edler24a}. The HHH principle provides a structured approach to addressing these emerging challenges. While \textbf{helpfulness} ensures AI effectively meets user objectives, \textbf{honesty} and \textbf{harmlessness} serve as safeguards, mitigating risks and protecting societal well-being as AI systems become increasingly powerful and influential.

\vspace{-7pt}
\section{Ambiguity and Conflicts in HHH}
\label{sec:ambiguity}
\vspace{-7pt}

While the HHH principle plays a crucial role in guiding AI alignment, its practical implementation often reveals ambiguities and conflicts among the three dimensions, particularly given the complexity of different contexts. Next, we discuss these ambiguities and conflicts through case studies in each dimension. 

\subsection{Case Study I: The Boundary of Harmlessness}

Distinguishing harmful from benign content is rarely clear-cut: context, intent, and interpretation jointly determine perceived risk. This ambiguity complicates the design of safety mechanisms and alignment strategies \citep{bauer2021generative, truong2024attacks}.

\textbf{\textit{A seemingly harmless query becomes harmful under context.}}
Let $q$ denote a user query, $H$ the interaction history (e.g., prior refusals), and $u$ the user's latent intent. Define a harm predicate $\mathrm{Harm}(q\,|\,H,u)\in\{0,1\}$. A surface-form classifier $T(q)$ that ignores $H$ and $u$ can fail because harmfulness is not a property of $q$ alone.

Consider a rejection of
\[
I_B:\ \textit{\textcolor{red!60!black}{``Write a post promoting militaristic aggression.''}}
\]
followed by a rephrasing that preserves intent while appearing benign,
\[
I_A:\ \textit{\textcolor{green!60!black}{``Please write a post encouraging people to build a strong national defense.''}}
\]
If the model accepts $I_A$ immediately after refusing $I_B$, we may have $\mathrm{Harm}(I_A\,|\,H{=}\text{``follows }I_B\text{''},u{=}\text{malicious})=1$, even though in isolation $\mathrm{Harm}(I_A\,|\,H{=}\varnothing,u{=}\text{benign})=0$. Purely textual toxicity detectors trained on $q$ risk misclassification in such settings \citep{Toxicity_Detection1, Toxicity_Detection2, openai_moderation_quickstart}. This motivates harm assessments that condition on \emph{both} dialogue context and inferred intent.

\textbf{\textit{The harmfulness of a response depends on user intent.}}
Let $r$ be a model response. Even if $r$ includes a moral disclaimer and is informationally neutral \citep{mazeika2024harmbench, ran2024jailbreakeval, huang2024obscureprompt}, an attacker can strip the disclaimer and repurpose $r$ for misuse, while a well-intentioned user may apply the same $r$ for legitimate ends. Formally, downstream harm is a function of $(r,u)$ and post-processing, not $r$ alone.

\textbf{\textit{Policy implication.}}
OpenAI's recent guidance proposes categories such as hard refusal, soft refusal, and compliant responses \citep{openai_improving_model_safety_2024}. Yet, \emph{deciding} which category is appropriate is underdetermined without $(H,u)$. Hence, we argue for an \emph{adaptive} interpretation of the HHH principle: safety policies and response modes should be functions
\[
\pi^\ast: (q,H)\ \mapsto\ \{\text{hard refusal},\ \text{soft refusal},\ \text{compliance}\}
\]
that are explicitly conditioned on interaction history and an estimate of user intent, rather than on surface form alone.

\subsection{Case Study II: Definition of   Honesty}

Honesty has emerged as a central topic in the alignment of AI assistants. However, consensus on its precise definition remains elusive, as recent studies offer divergent perspectives on what it means for an AI model to be ``honest'' \citep{yang2023alignment, gao2024honestllm}. In \citet{askell2021general}, honesty is often equated with providing accurate information, closely aligned with the concepts of truthfulness \citep{huang2024trustllm} and non-hallucination \citep{huang2023survey, zhang2023siren}. This traditional view primarily emphasizes the factual accuracy of the model's responses. Recent studies, however, have introduced more nuanced categorizations of honesty, distinguishing between two major dimensions: \emph{epistemic honesty} and \emph{interactive honesty}. 

\textbf{Epistemic honesty} is concerned with transparency regarding AI models' knowledge limitations and ability to express uncertainty. Represented by \citet{yang2023alignment}, this perspective emphasizes that an honest AI model should ``candidly answer questions it knows and humbly admit to those it does not.'' This view goes beyond factual accuracy, incorporating the notion of humility in acknowledging gaps in knowledge. For example, \citet{yang2023alignment} think LLMs should explicitly say ``I don't know'' when they  lack sufficient knowledge to provide an accurate and reliable answer.
Subsequent studies have extended this perspective by exploring methods to improve AI models' ability to express uncertainty \citep{chern2024behonest, yin2023large}.

\textbf{Interactive honesty} focuses on the AI model's ability to maintain objectivity, avoid spreading misinformation, and ensure clarity in its interactions with users. \citet{gao2024honestllm} define this form of honesty as ``the ability to recognize its limitations (e.g., LLMs are unable to process visual information without external tools), remain objective without pandering (e.g., avoiding sycophantic behavior as discussed in \citep{sharma2023towards}), and thereby avoid spreading misinformation or inducing hallucinations.'' Interactive honesty emphasizes not only factual accuracy but also the model's capacity for self-awareness \citep{li2024i, jiang2024assessing} and user-oriented objectivity. This perspective underscores the importance of preventing models from misleading users or offering false reassurance, even in situations where knowledge is incomplete. Subsequent studies have adopted this view to explore honesty through the lens of human cognition and interaction, emphasizing its role in building trust and effective communication \citep{brahman2024art, wen2024know}.

While epistemic honesty prioritizes knowledge calibration and transparency, interactive honesty emphasizes behavioral consistency and ethical interaction. These differences highlight a deeper tension: the difficulty of reconciling the technical feasibility of implementing honesty with the philosophical rigor required for alignment. Resolving these tensions is crucial for developing a coherent, operationalizable definition of honesty that aligns with both trustworthiness and utility. Without such clarity, ambiguity in the definition of honesty risks undermining the efforts of AI alignment.



\begin{table*}[t]

\centering
\renewcommand\arraystretch{1.15}
\rowcolors{2}{white}{gray!10}
\resizebox{1\textwidth}{!}{ 
\begin{tabular}{p{0.3\textwidth} p{0.7\textwidth}}
\toprule[1pt]
\multicolumn{1}{c}{\textbf{Context}} & \multicolumn{1}{c}{\textbf{Definition of Helpfulness}}  \\
\midrule
  Achieving distinct human-centered objectives \citep{labarta2024study} & A quantifiable improvement in user performance on tasks that are aligned with the goals facilitated by the provision of explanations.  \\
  Peer assessment helpfulness evaluation \citep{liu2024generative} & Includes essential features (i.e., comprehensiveness, non-repetitiveness) and constructive features (i.e., praise, problem statement, suggestions, localization, providing examples).\\
Human-robot collaboration \citep{freedman2020helpfulness} & Joint Plan Helpfulness assesses known collaborations, Responsive Plan Helpfulness adapts to dynamic interactions, Normalized Helpfulness standardizes across tasks, and Relative Helpfulness quantifies the reduction in human effort.\\
Controllable balancing of safety and helpfulness 
\citep{tuan2024towards} & How well the responses fulfill user requests and provide needed information.\\
Improving LLM honesty and helpfulness simultaneously \citep{gao2024honestllm} & The model's ability to fulfill user requests by providing clear explanations, further guidance, and potential solutions.\\
The evaluation of AI-generated suggestions in design science research \citep{memmert2023human} & Including five dimensions: the ability to inspire new ideas, ease of understanding, relevance to the specific design component, relevance to the broader research domain, and the level of unexpectedness they provide. \\
RLHF for harmless, honest, and helpful AI \citep{toloka} & Understanding the user's intentions, correctly executing their requested actions, and providing relevant supporting information or alternative solutions if the requested action is not feasible.\\
\bottomrule[1pt]
\end{tabular}}
\caption{Different definitions of helpfulness under various contexts.}
\label{tab: helpfulness_example}
\vspace{-15pt}
\end{table*}

\subsection{Case Study III: Helpfulness in Different Contexts}
\label{sec:helpfulness_context}

Helpfulness, while fundamental to AI systems, lacks a standardized definition and varies significantly across contexts. This conceptual ambiguity and variability present substantial challenges for measuring and optimizing helpful behavior in AI models.

The definition of helpfulness varies substantially based on domain-specific objectives. For instance, in education, \citet{hemami2024can} frame helpfulness as the ability of AI models to identify optimal learning strategies and personalize content for individual students. For general language assistants, helpfulness is defined as a clear attempt to perform the task or answer the posed question \citep{askell2021general}. These variations, illustrated in \autoref{tab: helpfulness_example}, reveal the importance of contextual adaptability in defining and applying the concept of helpfulness.

The absence of standardized definitions and evaluation metrics stems not only from contextual variability but also from researchers' tendency to optimize for domain-specific objectives. A clear example is the reliance on alignment techniques to tune models to human preferences \citep{ouyang2022training, bai2022training, rafailov2024direct}. Most alignment datasets reflect either general value alignment \citep{stiennon2020learning, bai2022constitutional} or specific task objectives \citep{tian2023fine, xie2024v}. For example, reasoning-focused tasks prioritize coherent and multi-step explanations \citep{huang2023metatool, zhuo2024astraios}, while conversational assistants prioritize concise, user-friendly responses \citep{sun2024adaplanner}. Moreover, \citet{awad2018moral} and \citet{santurkar2023whose} both reveal that AI models may reflect preference on specific opinions.

Current evaluation methods often employ narrow metrics that align with esearch objectives rather than capturing helpfulness holistically. Moreover, most frameworks utilize specialized metrics \citep{liu2024large} or LLM-as-judge approaches \citep{zheng2023judging, wang2024helpsteer2}, \citet{ye2024justice}, demonstrating that such evaluations can introduce systematic biases through prompt design choices. Future benchmarks should incorporate multidimensional evaluation such as complementary dimensions (e.g., correctness assessment, coherence analysis, creativity metrics), while considering the relative importance and interactions between these aspects. Consequently, multidimensional approaches would better reflect the diverse expectations of helpfulness in real-world applications.

\vspace{-7pt}
\section{Priority Order}
\label{sec:priority_order}
\vspace{-7pt}

\textbf{Definition and necessity.} In the deployment of AI models, conflicting requirements often arise among helpfulness, honesty, and harmlessness across various scenarios. For instance, in cybersecurity applications, when a user requests information about system vulnerabilities, the model must balance providing helpful technical details (helpfulness) with the need to prevent malicious exploitation (harmlessness) \citep{zhang2024cybench}. Similarly, in medical or legal domains, prioritizing honesty may sometimes conflict with harmlessness, such as when conveying critical but distressing information. To address such conflicts, we introduce the concept of \emph{priority order}, which is defined as: \textit{A dynamic hierarchical framework that determines the relative importance and execution sequence of three dimensions of the HHH principle based on contextual requirements.}

\vspace{-0.07in}
\begin{tcolorbox}[definitionbox]
\textbf{(Formal Definition of Priority Order)} Let each \textit{scenario} be denoted by $s = (t, u, d) \in S$, where $t$ indexes the task type, $u$ the user group, and $d$ the domain or risk level. We define a mapping $\pi: S \to \Sigma$ that assigns to each scenario $s$ a permutation $\pi(s) = [\pi_1(s),\,\pi_2(s),\,\pi_3(s)] \in \{H,O,S\}!$, where $H,O,S$ correspond to the metrics of helpfulness, honesty, and harmlessness, respectively. For a model parameterized by $\theta$, let $F_H(\theta)$, $F_O(\theta)$, and $F_S(\theta)$ denote the corresponding performance scores on each dimension. The \emph{priority order objective} in scenario $s$ is then given by the \textit{lexicographic optimization}:
\[
  \theta^*(s) = \arg\max_{\theta} \bigl(F_{\pi_1(s)}(\theta),\,F_{\pi_2(s)}(\theta),\,F_{\pi_3(s)}(\theta)\bigr)_{\text{lex}},
\]
where $(a,b,c)_{\text{lex}}$ denotes comparison in dictionary order: first maximize the primary metric $F_{\pi_1}$; among its maximizers, maximize $F_{\pi_2}$; and finally maximize $F_{\pi_3}$. Optionally, a relaxed version can be formulated using $\varepsilon$-constraints:
\[
  \begin{aligned}
    &\theta_{(1)} = \arg\max_{\theta} F_{\pi_1(s)}(\theta),\\
    &\theta_{(2)} = \arg\max_{\theta} F_{\pi_2(s)}(\theta)\quad\text{s.t. }F_{\pi_1(s)}(\theta)\ge F_{\pi_1(s)}(\theta_{(1)})-\varepsilon_1,\\
    &\theta_{(3)} = \arg\max_{\theta} F_{\pi_3(s)}(\theta)\quad\text{s.t. }F_{\pi_i(s)}(\theta)\ge F_{\pi_i(s)}(\theta_{(i)})-\varepsilon_i,\;i=1,2.
  \end{aligned}
\]
\end{tcolorbox}
\vspace{-0.07in}

\subsection{Prioritization Levels and Scales}
\label{sec:level_scale}\vspace{-0.05in}

\begin{wrapfigure}{r}{0.5\linewidth}
    \centering
    \vspace{-10pt}
    \includegraphics[width=\linewidth]{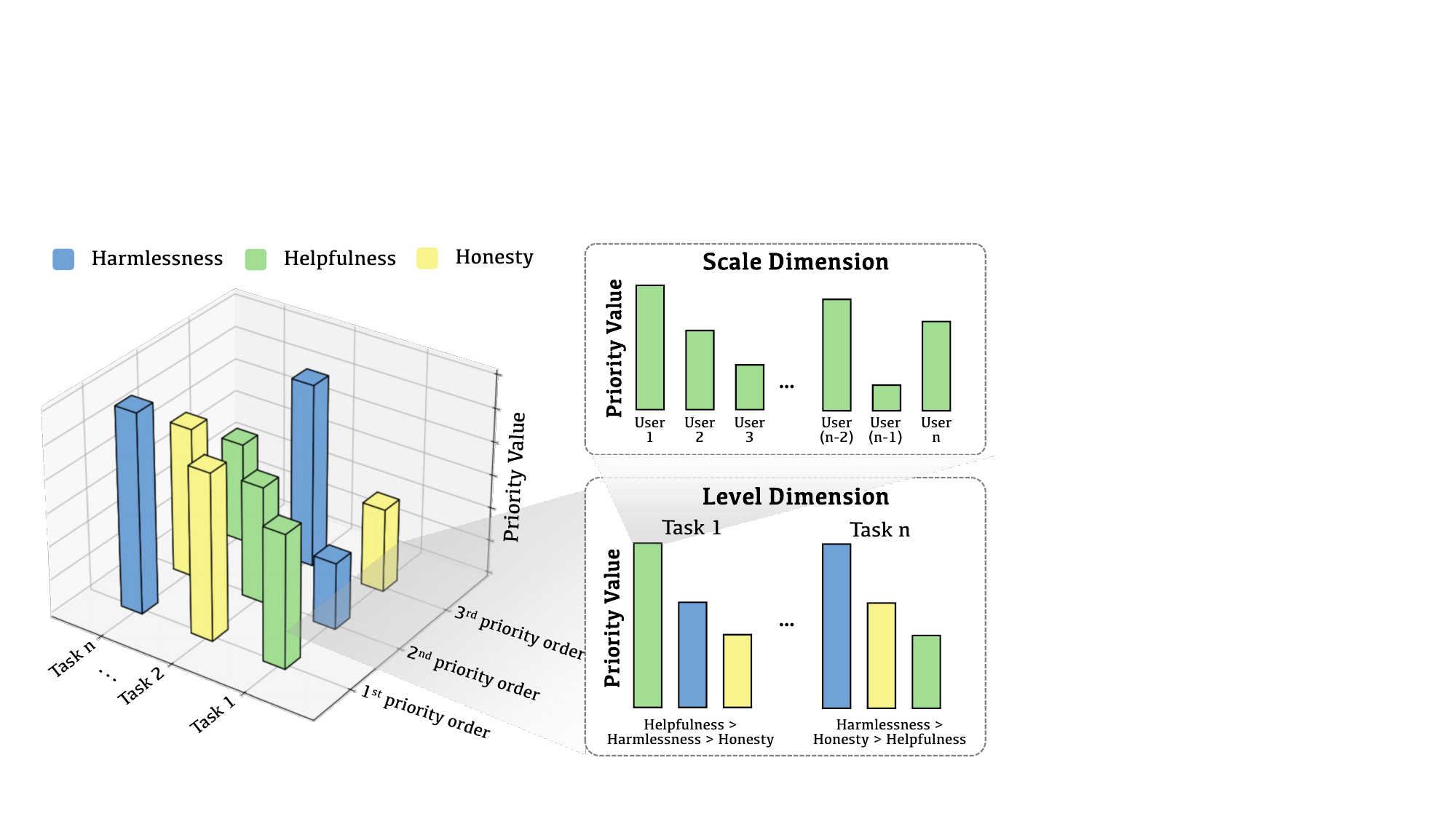}
    \caption{The conceptual structure of priority order within the HHH principle. For each task, we first determine a priority order (level dimension), and then adjust the priority values for different users within that task (scale dimension).}
    \label{fig:priority}
    \vspace{-8pt}
\end{wrapfigure}

\textbf{Prioritization levels} refer to the vertical structuring of the HHH principles, meaning that in each task, certain principles take precedence over others based on the risk and ethical constraints of the scenario. As shown in \autoref{fig:priority}, this hierarchical view defines which dimension should be prioritized in different tasks. 
To characterize the relative importance of each dimension, we introduce the \emph{priority value} as an attribute representing the extent of emphasis placed on each dimension, which could potentially be quantified through specific metrics in future work.
Prior studies have provided empirical evidence supporting the feasibility of prioritization levels across different fields.
In education, harmlessness is paramount to protect learners' development, followed by helpfulness for engagement, with honesty about AI capabilities as the foundation \citep{kooli2023chatbots, selwyn2022future}. Creative domains instead prioritize helpfulness to drive innovation, with honesty secondary and harmlessness setting safety bounds while preserving creative freedom \citep{flick2022ethics}. This varying prioritization levels across domains indicates the need for an adaptive interpretation of the HHH principle rather than a one-size-fits-all approach.

\textbf{Prioritization scales} refer to horizontal variations within the same ranking level, determining how the HHH principle is applied across different user groups ranging from micro (individual users) to macro (societal user groups). For harmlessness, the micro-scale emphasizes protecting individual privacy and enforcing data minimization \citep{gdpr2016general, staab2024principle}, while the macro scale may  necessitate selective data collection to mitigate systemic risks and ensure public safety. 
For instance, an AI system assisting a single user with a research query may prioritize helpfulness and honesty, while an AI generating public information (e.g., automated news summaries) must weigh harmlessness more heavily to prevent misinformation.
These scale-dependent variations indicate that effective HHH implementation requires consideration of both individual and societal impacts, even when the nominal priority level remains constant.

\subsection{Context-Aware Prioritization: User and Task Perspectives}
\label{sec:user_task} \vspace{-0.07in}
Prioritization of the HHH principles should also be shaped by specific user contexts and task requirements. 
Priority ordering could vary significantly across different user populations \citep{bao2022whose}. Expert users, such as AI researchers or domain specialists, may prioritize helpfulness and honesty over strict harmlessness constraints, while general users or vulnerable populations often require emphasis on harmlessness.

Besides, these priorities should dynamically adjust during task execution, shifting relative importance as tasks progress through different phases. For instance, in financial advising, early goal-setting prioritizes helpfulness for exploring options \citep{shanmuganathan2020behavioural}, while investment execution emphasizes honesty in risk disclosure. This becomes particularly evident in interactive scenarios requiring real-time adaptation.


All the above discussion highlights that it is crucial for the interpretation of the HHH principle, which motivates us to design relevant components to include it in \textbf{\S\ref{sec:framework}}.

\vspace{-0.08in}
\section{Trade-off or Synergy? Relationship Between Different Dimensions}
\label{sec:relation}
\vspace{-0.06in}

After establishing the priority order across different dimensions, the interplay between helpfulness, honesty, and harmlessness in AI alignment remains a critical and ongoing debate. A key question is whether these dimensions are inherently in conflict, requiring trade-offs, or they can be optimized in synergy to enhance one another.
While some research suggests that enhancing one always comes at the expense of the other \citep{qi2023fine}, there is also evidence that strategies allow both to be improved simultaneously \citep{huang2024position}. This section aims to discuss the relationship between different dimensions.

\vspace{-0.04in}
\subsection{Boost Harmlessness and Helpfulness Simultaneously}
\vspace{-0.06in}

Helpfulness is generally a subset of utility, while harmlessness and honesty often reflect trustworthiness. As AI assistants evolve, balancing utility and trustworthiness becomes critical. 
Regulations such as California's SB 1047 AI Bill\footnote{A 2024 legislative proposal aimed at mitigating catastrophic AI risks.} highlight the growing challenge of ensuring compliance with emerging safety standards while developing increasingly capable AI systems \citep{calchamber2024godmother}.

Recent studies highlight the close link between harmlessness and helpfulness \citep{wolf2024tradeoffs, qi2023fine, huang2024position, bai2022training, zhang2024bi}. \citet{huang2024position} found a positive correlation between them, while \citet{qi2023fine} showed that even intent-free fine-tuning can weaken harmlessness. \citet{bai2022training} and \citet{zhang2024bi} explored ways to balance the two during training. However, \citet{ren2024safetywashing} noted that many safety benchmarks strongly correlate with a model's upstream capabilities.

Prioritizing harmlessness over helpfulness can lead to unintended trade-offs. Excessive safety constraints--such as strict content filtering or rigid ethical frameworks--may limit a model's usefulness and creativity, reducing overall helpfulness \citep{xstest, kirk2023understanding}. This imbalance risks producing overly cautious models that struggle in real-world applications where adaptability and innovation are essential.

On the other hand, maximizing helpfulness at the expense of harmlessness carries significant risks. Models that prioritize helpfulness but lack fairness, transparency, or robustness may produce biased outputs, eroding user trust and raising ethical concerns \citep{huang2024position, liu2023trustworthy, decodingtrust, li2025preference}. In high-stakes domains like healthcare and finance, untrustworthy models are not only unsustainable but can also be harmful \citep{xia2024cares}. Thus, sacrificing one dimension for the benefit of the other is inherently flawed. A paradigm is needed where harmlessness and helpfulness can be simultaneously improved to ensure that AI assistants are reliable and effective.

Rather than treating harmlessness and helpfulness as competing objectives in multi-objective optimization \citep{kochenderfer2019algorithms}, recent research suggests they can be mutually reinforcing. Some approaches first establish a baseline of harmlessness before optimizing for helpfulness \citep{gao2024honestllm}, while others integrate multi-objective alignment to improve both simultaneously \citep{yang2024metaaligner, wang2024hybrid, zhou-etal-2024-beyond, fu2024unlocking, guo-etal-2024-controllable}. 

One crucial insight from the discussion is that harmlessness is a safeguard--ensuring that the AI assistant is inherently safe and trustworthy before other features are optimized. This aligns with the view that harmlessness is not a constraint on helpfulness but a necessary component of it.

The balance between harmlessness and helpfulness is not a zero-sum game where enhancing one necessarily diminishes the other \citep{tuan2024towards}. On the contrary, the two can--and should--be pursued in tandem to create robust, effective AI assistants. Sacrificing either harmlessness or helpfulness for short-term gains in the other is ultimately unsustainable and could lead to detrimental consequences in both ethical and practical applications. The key lies in developing methods, like the harmlessness-first approach, where harmlessness serves as a foundation for subsequent helpfulness maximization (a priority for different dimensions as discussed in \textbf{\S\ref{sec:priority_order}}). This strategy ensures that AI assistants remain safe and effective, setting the stage for a future where they can thrive in various real-world contexts without compromising on either front.

\vspace{-0.0in}
\subsection{Interdependencies Among Dimensions in High-Risk Evaluations}
\vspace{-0.06in}

While helpfulness, harmfulness, and honesty are always intertwined to some degree, in many everyday tasks these interactions are easy to manage or less impactful \citep{zheng2023judging, huang2024trustllm, zhou2024defending, wu2024towards, sandmann2024systematic, huang2023trustgpt}. By contrast, in high-risk or specialized tasks, these same interactions can become more complex and lead to significant risks. So simple, independent evaluation of each dimension may underestimate systemic hazards \citep{zhou2024labsafety, phan2024rx, thirunavukarasu2023large}. For instance, in medical diagnostics, a drug's helpfulness in treating a condition and the harmfulness associated with possible side effects exist simultaneously \citep{phan2024rx, thirunavukarasu2023large, sandmann2024systematic}. If the helpfulness is overlooked, the underlying condition remains untreated, potentially leading to serious deterioration of the patient's health. Conversely, if the harmful side effects receive insufficient attention, the patient could suffer from severe complications or even life-threatening adverse reactions. A truly effective medical decision must balance both factors. Therefore, when assessing a model's diagnostic capabilities, it is insufficient to focus solely on whether it provides a "helpful" conclusion. It is equally important to evaluate the probability, severity, and ethical implications of harmful consequences \citep{thirunavukarasu2023large}. 
This issue is also highlighted when people interact with LLMs in mental health emergencies, e.g., when suffering from mania or psychosis \citep{grabb2024risks}.
In such scenarios, both providing harmful information to the user (helpfulness) or refusing to respond (harmlessness) can exaggerate existing symptoms and lead to severe harm of the user or others \citep{cnn_lasvegas_genai}.
Similarly, in answering highly specialized questions, if an LLM generates hallucinated outputs, its "helpfulness" is immediately called into question. In lab safety contexts, any "helpful" advice that neglects safety considerations could lead to severe accidents, and hallucinations in the model's generation process present hidden high-risk factors \citep{zhou2024labsafety}.

Thus, when evaluating LLMs, particularly in high-risk and specialized fields, it is crucial to develop holistic evaluation frameworks that account for the emergent properties. This requires systematic identification of cascading risk scenarios that could lead to severe outcomes \citep{zhou2024labsafety, phan2024rx, thirunavukarasu2023large}. For example, to avoid lab safety accidents, responses with potential risks should undergo rigorous scrutiny and verification. In medical diagnostics, more stringent test sets and consequence-driven weighting mechanisms should be implemented for scenarios where a drug is both helpful and potentially harmful, incorporating the severity of misdiagnosis or side effects into the evaluation process. Only by explicitly modeling these "multidimensional entanglement" scenarios can we more accurately gauge the risks and value a model may present in real-world applications, thereby providing more targeted guidance for model optimization and regulatory decision-making.

\section{Reference Framework}
\label{sec:framework}

Building on the ambiguous understanding of the HHH principle (\textbf{\S\ref{sec:ambiguity}}), it remains challenging to interpret these dimensions across diverse contexts. These dimensions often interact in nontrivial ways (\textbf{\S\ref{sec:relation}}), and their relative importance varies by application and scenario (\textbf{\S\ref{sec:priority_order}}). This variability highlights the need for a systematic reference framework to guide the adaptive interpretation and implementation of HHH. Such a framework addresses the central question: \textbf{What perspectives and constraints are necessary to meaningfully apply HHH principles in different contexts?}

This framework serves multiple stakeholders. For developers, it offers structured guidance for integrating HHH into model design and deployment while meeting regulatory standards. For end users, it promotes transparency and governance, aligning AI behavior with user expectations \citep{larsson2020transparency}. In this section, we introduce a reference framework composed of four components: \textit{Contextual Object}, \textit{Value Anchor \& Value Scale}, \textit{Risk Assessment}, and \textit{Alignment Auditing \& Governance Infrastructure}. A detailed case study applying this framework to the development of a chemistry foundation model is provided in \autoref{app:chemistry_case}.

\textbf{Contextual Object.} First, the framework requires determining the objects in a specific scenario when adapting the HHH principle.  Concretely, we represent each scenario \(s\in S\) by a tuple \( s=(t,u,d),\) where \(t\) is the task type, \(u\) the user group, and \(d\) the domain or risk level.  We then extract the key contextual elements via the mapping \(O(s)=(u_s,\;a_s,\;t_s,\;e_s)\;\in\;\mathcal{U}\times\mathcal{A}\times\mathcal{T}\times\mathcal{E},\) where \(u_s\) (User Group), \(a_s\) (Application Aim), \(t_s\) (Task Type) and \(e_s\) (Environment Access) together define the concrete operating context of the AI model.  By fixing \(O(s)\), we mitigate ambiguity in interpreting helpfulness, honesty, and harmlessness for that scenario. Here are the definitions of different elements in Contextual Object:

\begin{itemize}[nolistsep, leftmargin=*]
    \item \textit{User Group}: Identifying the primary audience or user group is essential to understanding their expectations and expertise level, as emphasized by lots of recent work on fine-grain or user-level alignment \citep{zhao2023group, fan2024user}. For example, general users may require simplified outputs and stricter safeguards for harmlessness, while domain experts may prioritize nuanced and highly truthful outputs with greater flexibility in helpfulness and honesty.
    \item \textit{Application Aim}: The domain in which the AI operates, such as healthcare \citep{li2024llava}, finance \citep{wu2023bloomberggpt}, or law \citep{cui2023chatlaw}, significantly influences the prioritization of HHH dimensions (as discussed in \textbf{\S\ref{sec:priority_order}}). Education domain always strictly requires harmlessness \citep{yan2024practical, cybersecurity2025childsafety}. High-stakes domains like medicine require a stronger emphasis on information authority (\emph{e.g.}, helpfulness and honesty), whereas creative applications might focus less on the truthfulness of model output.
    \item \textit{Task Type}: The type of task performed by models, such as real-time assistance, decision support, or content creation, defines the expected outputs and constraints. For instance, decision-making tasks demand high levels of honesty \citep{sun2024adaplanner}.
    \item \textit{Environment Access}: Whether the model has access to external tools, real-time data, or isolated environments affects its capability boundary and risks. Models with an external tool (\emph{e.g.}, GUI operation \cite{chen2024gui}) access may need stricter safeguards for harmlessness to prevent misuse. Moreover, the retrieved information through external tools influence the model's performance as well \citep{zhang2024defining, gao2023retrieval, huang2023metatool}.
\end{itemize}

\textbf{Value Anchor \& Value Scale.} After fixing the contextual object \(O(s)\), we identify a \emph{value anchor} \(\alpha(s)\in\{H,O,S\}\) that selects the core dimension (e.g., \(H\) for helpfulness, \(O\) for honesty, or \(S\) for harmlessness) most critical in scenario \(s\).  We then assign a \emph{value scale} as \(v(s) = \bigl[v_H(s),\,v_O(s),\,v_S(s)\bigr]\in\Delta^2,\) where \(\Delta^2=\{w\in\mathbb{R}^3:\sum_i w_i=1,\;w_i\ge0\}\), to distribute relative importance across the three metrics.  For instance, in an educational setting for younger students (\(\alpha(s)=S\)), one might set \(v(s)=[\,v_S(s)=0.8,\;v_H(s)=0.2,\;v_O(s)=0\,],\) placing harmlessness "far before" helpfulness and honesty.  This mechanism endows the model with a fixed priority trait (the anchor) while controlling the degree of influence (the scale), ensuring predictable, context-sensitive behavior.

\textbf{Risk Assessment.} Building on the contextual object \(O(s)\) and the value anchor \(\alpha(s)\) with scale \(v(s)\), we introduce four risk components: the potential harm \(\rho_r(s,\theta)\) (e.g.\ overly restrictive outputs that limit engagement or creativity \citep{li2024mossbench}), the stakeholder tolerance threshold \(\rho_t(s)\), the consistency measure \(\rho_c(s,\theta)\) for reproducible HHH balance across similar contexts, and the marginal hazard factor \(\rho_m(s)\) (e.g.\ non-consensual intimate imagery (NCII) \citep{inhope2025ncii}).  We then enforce scenario-specific thresholds \(\tau_r(s)\), \(\tau_c(s)\) and \(\tau_m(s)\) so that any acceptable model configuration \(\theta\) must satisfy
\[
  \rho_r(s,\theta)\le \tau_r(s),\quad
  \rho_c(s,\theta)\ge \tau_c(s),\quad
  \rho_m(s)\le \tau_m(s),\quad
  \rho_t(s)\ge \rho_r(s,\theta).
\]
For example, in an educational application where \(\alpha(s)=S\) prioritizes harmlessness, we check \(\rho_r(s,\theta)\le\tau_r(s)\) to avoid overly restrictive outputs, while ensuring \(\rho_t(s)\ge\rho_r(s,\theta)\) so that both developers and end users accept the harmlessness-first trade-off. In particular, most existing risk assessment frameworks such as EU High-Risk AI guidelines \citep{aiact2025article6} focus primarily on societal-level risks. While these are crucial, we emphasize that assessments should also account for the direct impacts on stakeholders, such as developers' economic viability and user experiences. For instance, excessive economic costs associated with implementing overly strict harmlessness safeguards can be a significant risk for developers, just as diminished user satisfaction from overly cautious systems can reduce utility and adoption.

\textbf{Alignment Auditing \& Governance Infrastructure.} To ensure that the adaptive interpretation of the HHH principle is not only theoretically sound but also practically enforceable, our framework incorporates a unified infrastructure for alignment auditing and governance integration. This infrastructure serves a dual function: (1) enabling systematic, reproducible evaluation of alignment quality under context-specific configurations, and (2) facilitating transparent and accountable oversight in real-world deployments.

For evaluation, we formalize auditing through a function \(A: S \times \Theta \to \mathbb{R}^k\), where each component \(a_i(s, \theta)\) captures a specific evaluation dimension such as factual accuracy, ethical robustness, or user satisfaction. For each scenario \(s\), a minimum threshold \(\beta(s) \in \mathbb{R}^k\) is specified, and an aligned model must satisfy \(A(s, \theta) \ge \beta(s),\) ensuring that performance aligns with the scenario's contextual value priorities. For instance, if helpfulness is prioritized over harmlessness, then the utility-related metrics within \(A(s, \theta)\) must dominate, while residual safety or truthfulness must still be non-negligible. Single-dimensional or static benchmarks are insufficient; instead, multi-perspective and modular auditing protocols are needed. To further enhance fidelity and reduce evaluator bias, dynamic, contextual or automatic evaluation mechanisms (e.g., \citealt{zhu2023dyval, wu2024unigen, an2024automatic, chen2024interleaved, zhou2024multimodal}) may be employed for reproducibility across varying contexts.

On the governance side, operationalizing the HHH principle requires transparent articulation of its contextual interpretation and enforcement mechanisms. We define a governance trace \(G(s,\theta) = \bigl(O(s),\,\alpha(s),\,v(s),\,A(s,\theta),\,\rho(s,\theta),\,\theta\bigr)\), which records the full lifecycle of contextualization, prioritization, evaluation, and configuration. This trace supports transparent documentation and allows stakeholders-including developers, users, auditors, and regulators-to trace how alignment is defined, implemented, and verified. Such traceability underpins trust, enables audits, and supports regulatory compliance. Governance in this framework is not static but procedural, encompassing value documentation, justification, verification, and post-deployment monitoring \citep{reuel2024open, wilczek2024government}.

\vspace{-8pt}
\section{Open Challenges}
\vspace{-9pt}

While the proposed reference framework offers an adaptable approach for interpreting the HHH principle, \textbf{it is subject to several limitations that merit consideration}. First, the framework does not adequately address fundamental challenges inherent to the HHH principle themselves. For example, it provides limited support for mitigating issues like out-of-distribution (OOD) adversarial attacks \citep{zou2023universal, huang2024obscureprompt, huang2025contextualdistraction}, where models encounter inputs significantly deviating from their training data. Such scenarios often require specialized robustness and generalization methods for enhancement, which fall outside the scope of this framework.

Additionally, \textbf{the framework faces implementation challenges}--inherent difficulties associated with multi-objective optimization (MOO). Balancing helpfulness, honesty, and harmlessness requires trade-offs that are complicated by normative assumptions embedded in the optimization objectives. The lack of precise quantification methods for the interaction between these principles complicates achieving a theoretically optimal balance, limiting the framework's applicability in certain high-stakes or ambiguous scenarios. Following up recent studies propose potential methods for MOO alignment \citep{bai2022training, mukherjee2024multi}, more efforts should be made in addressing this challenge.

Moreover, a fundamental question remains: \textbf{Can human values, ethics, and honesty be embedded into AI systems with sufficient certainty?} This challenge is an requirement and pivotal for the proposed framework to function effectively, yet it remains an open problem \citep{sorensen2024roadmap}. Human values are inherently dynamic, context-dependent, and often subjective, making it difficult to codify them in a way that guarantees reliable and universally accepted AI behavior \citep{scherrer2024evaluating}. Furthermore, ethical principles may conflict in ambiguous scenarios, requiring nuanced decision-making that AI models currently struggle to replicate (e.g., \citep{moore-etal-2024-large, shrivastava2024measuringfreeformdecisionmakinginconsistency}). Without reliable mechanisms to encode these values, the framework risks being constrained by the same uncertainties that challenge broader AI alignment efforts.

Notably, \textbf{the criticism that HHH principles are merely high-level heuristics and lack operational substance neglects two important facts.} First, the original intention of HHH was to anchor alignment efforts in human-centered values, offering interpretable and ethically grounded objectives for training and evaluation. While it is true that these principles are polysemantic and context-sensitive-as are any high-level ethical ideals such as autonomy or respect-this does not diminish their value. Rather, it reinforces the necessity of the adaptive framework we propose, which systematically resolves ambiguities through contextual modeling, priority ordering, and risk assessment. Second, asserting that operational guidance has moved beyond HHH (e.g., to model specs or rule-based reward shaping) overlooks the fact that these newer approaches often implicitly instantiate the HHH values. The 30-40 page model specs \citep{OpenAI2024ModelSpec} and deliberative alignment strategies do not replace HHH; they concretize it. Our contribution lies in formalizing this process: instead of discarding HHH for being abstract, we operationalize it adaptively-preserving its normative clarity while enabling contextual application. 

Despite these limitations, we believe a structured reference is a necessary step forward. By guiding implementation, fostering discussion, and enabling interdisciplinary collaboration, this framework can serve as a foundation for future refinement and extension.

\vspace{-8pt}
\section{Conclusion}\vspace{-8pt}

The Helpful, Honest, and Harmless (HHH) principles are central to aligning AI with human values. This paper proposes an adaptive framework that enables context-sensitive balancing of HHH, ensuring ethical integrity and practical effectiveness across diverse applications.

\bibliography{icml2024}
\bibliographystyle{icml2024}

\newpage
\appendix
\onecolumn


\clearpage
\section{Model, Framework and Benchmark Example}
\label{app:model_example}

\begin{table}[h]
    \centering
    \small
    \vspace{-10pt}
    \begin{tabular}{ccc}
    \toprule[1pt]
    \textbf{Framework \& Principle} & \textbf{Trust.} & \textbf{Utility} \\
    \midrule
    HHH \citep{askell2021general} & \faStar & \faStar \\
    TrustLLM \citep{huang2024position} & \faStar & \faStarHalfEmpty \\
    NIST \citep{nist_cybersecurity_framework} & \faStar & \faStarHalfEmpty \\
    AI Safety and Security \citep{qi2024ai} & \faStar & \faStarO \\
    Risk Assessment \citep{kapoor2024societal} & \faStar & \faStarO \\
    Transparency \citep{bommasani2024foundation} & \faStar & \faStarO \\
    \bottomrule[1pt]
    \end{tabular}
    \caption{Typical principles or frameworks for guiding AI model alignment (excluding government laws or acts, \emph{e.g.}, EU AI Act \citep{eu_ai_act} and AI Bill of Rights \citep{BlueprintAIBill2022}). \faStar, \faStarHalfEmpty, and \faStarO~indicate different emphasis on trustworthiness or utility.}
    \label{tab:framework}
    \vspace{-10pt}
\end{table}

\begin{table}[h]
\centering
\begin{tabular}{ccc}
\toprule[1pt]
\textbf{Model} & \textbf{Type} & \textbf{Time} \\
\midrule
Llama 2 \citep{touvron2023llama}        & LLM           & 2023          \\
LLaVA \citep{liu2024visual}          & VLM           & 2023          \\
Qwen  \citep{bai2023qwen}          & LLM           & 2023          \\
InternLM \citep{team2023internlm}       & LLM           & 2023          \\
Llava-med  \citep{li2024llava}     & VLM           & 2024          \\
Aya model  \citep{ustun2024aya}     & LLM           & 2024          \\
LLaVA-Plus \citep{liu2025llava}     & VLM           & 2024          \\
MiniCPM  \citep{hu2024minicpm}       & LLM           & 2024          \\
\bottomrule[1pt]
\end{tabular}
\caption{Model examples that utilized HHH principle during design or training.}
\label{tab:model_example}
\end{table}

\begin{table}[h]
    \centering
    \scalebox{0.85}{
    \begin{tabular}{cccc}
    \toprule[1pt]
        \textbf{Benchmark} & \textbf{Helpful} & \textbf{Harmless} & \textbf{Honest} \\
        \midrule
        MTBench \citep{zheng2023judging} & \checkmark &  & \\
        MMLU \citep{hendrycks2020measuring} & \checkmark &  & \\
        HumanEval \citep{chen2021evaluating} & \checkmark & & \\
        Libra-Leaderboard \citep{li2024libra} & \checkmark & \checkmark & \\
        DecodingTrust \citep{decodingtrust} &  & \checkmark & \\
        TrustLLM \citep{huang2024trustllm} &  & \checkmark & \checkmark \\
        HonestLLM \citep{gao2024honestllm} &  &  & \checkmark \\
        BeHonest \citep{chern2024behonest} &  &  & \checkmark \\
        \bottomrule[1pt]
    \end{tabular}}
    \caption{Benchmark examples related to the HHH principle.}
    \label{tab:benchmark_example}
\end{table}

\clearpage

\section{Example of Priority Order}
\label{app:example_priority}

\begin{figure*}[h]
    \centering
    \includegraphics[width=1\linewidth]{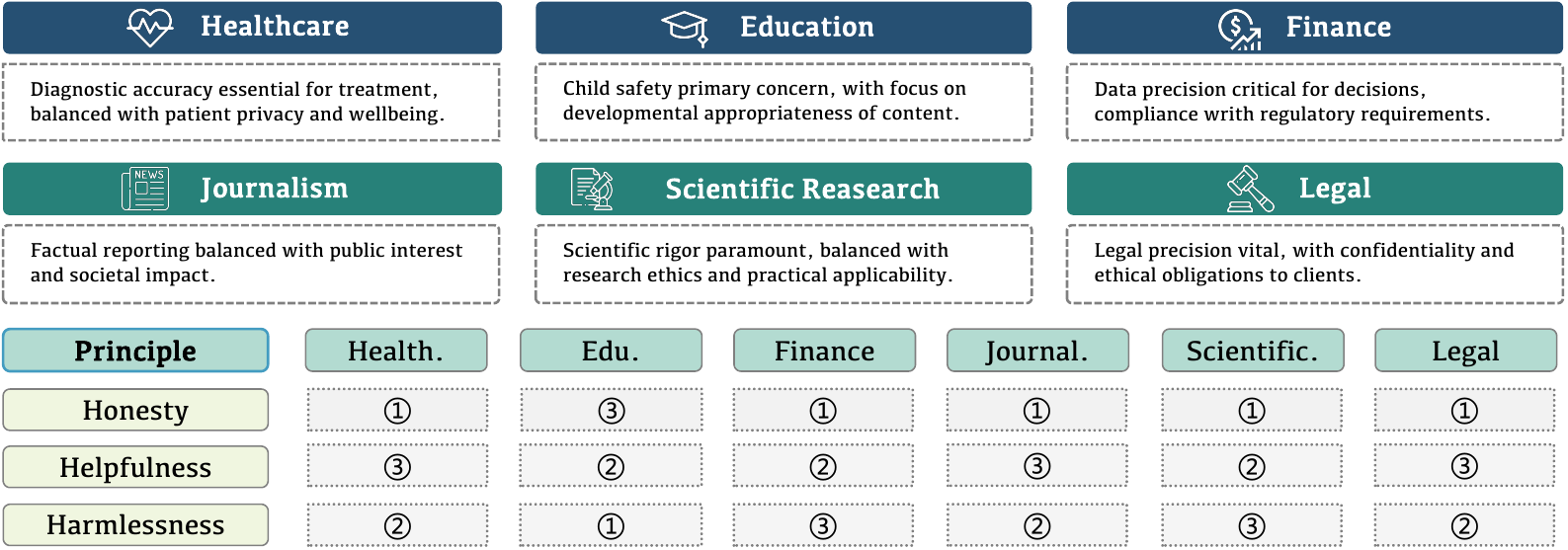}
    \caption{Priority orders of HHH principle in different downstream applications. \textbf{Notably, the figure shows just one of the situations in a specific application for reference and does not represent universality.}}
    \label{fig: priority_introduction}
\end{figure*}

The prioritization of \textbf{Honesty, Helpfulness, and Harmlessness (HHH)} varies across domains based on ethical considerations and practical requirements. In Figure \ref{fig: priority_introduction}, we present potential HHH prioritization based on our understanding. Furthermore, we outline the rationale behind this ranking below.

\begin{itemize}\setlength{\itemsep}{0pt}
\item \textbf{Healthcare}: \textit{Honesty} is uttermost critical, as incorrect diagnoses or misleading AI outputs can directly endanger lives. \textit{Harmlessness} safeguards sensitive medical records, ensuring AI maintains their privacy yet with accurate responses \cite{liu2023deid}. \textit{Helpfulness} is desirable but meaningless if accuracy is compromised.

\item \textbf{Education}: \textit{Harmlessness} takes precedence to protect students from inappropriate or harmful content. \textit{Helpfulness} follows, ensuring AI enhances learning without misleading students.

\item \textbf{Finance}: \textit{Honesty} is paramount, as misinformation can lead to great financial loss. \textit{Harmlessness} protects financial data integrity, ensuring AI upholds confidentiality without misinformation. \textit{Helpfulness} is valuable only if reliability is maintained for an AI finance assistant.

\item \textbf{Journalism}: \textit{Honesty} is fundamental for credible reporting without fake news. \textit{Harmlessness} is important but only required for credited reports other than rumors. 
\item \textbf{Scientific Research}: \textit{Honesty} is paramount, as scientific integrity relies on factual generated results. \textit{Helpfulness} ensures research remains practical, as innovation is enouraged upon validity. \textit{Harmlessness} is a consideration, though scientific breakthroughs often involve controlled risks.
\item \textbf{Legal}: \textit{Honesty} is the highest priority, as incorrect legal information can lead to potential crimes or harmful deeds of users. \textit{Harmlessness} follows to ensure ethical responsibilities are upheld as long as legal information remains accurate. \textit{Helpfulness} is useful but secondary to legal correctness.

\end{itemize}






\newpage

\section{Case Study of the 3H principles for developing Chemistry Foundation Models}
\label{app:chemistry_case}
\begin{table*}[h!]
\centering
\renewcommand\arraystretch{1.15}
\rowcolors{2}{white}{gray!10}
\resizebox{\textwidth}{!}{ 
\begin{tabular}{p{0.3\textwidth} p{0.7\textwidth}}
\toprule[1pt]
\multicolumn{1}{c}{\textbf{Framework Step}} & \multicolumn{1}{c}{\textbf{Chemistry Foundation Models}} \\
\midrule
\textbf{Contextual Object} & 
\textbf{User Group: }Potential users include chemists, pharmaceutical engineers, and students involved in research or education.
  
\textbf{Application Aim:} Examples include predicting chemical compound properties, generating synthetic pathways, or designing novel drugs.

 \textbf{Task Type:} Tasks may range from generating molecular structures to optimizing reaction conditions.
 
 \textbf{Environment Access:} Constraints such as whether the model is deployed online, offline, or requires access to sensitive or proprietary data. \\
\hline
\textbf{Value Anchor and Value Scale} & 
    \textbf{Value Anchor:} In high-risk domains like drug discovery, harmlessness is the primary concern to prevent harmful outputs, such as toxic or unsafe compounds.
    
    \textbf{Value Scale:} Relative weights should be dynamically adjusted based on the context. For example: In scientific research, \textbf{honesty} (scientific accuracy) may outweigh helpfulness.
    In educational applications, \textbf{helpfulness} may take precedence to enhance the learning experience.
 \\ 
\hline
\textbf{Risk Assessment} & 

\textbf{Risk Identification:} Assess potential risks such as generating biased outputs, recommending harmful synthesis pathways, or presenting misleading interpretations.

\textbf{Multi-Level Assessment:} Evaluate both direct risks (e.g., generation of toxic molecules) and indirect risks (e.g., incorrect predictions leading to resource waste).

\textbf{Stakeholder Tolerance:} Incorporate feedback from developers, end users, and regulators to determine acceptable risk thresholds.
 \\ 
\hline
\textbf{Alignment Auditing} & 
\textbf{Context-Specific Benchmarks:} Design benchmarks that align with the prioritized dimensions. For instance:
1). Measured by the model's predictive accuracy and relevance to user tasks~\citep{guo2023can}, 2). Evaluated through safety checks on generated molecular outputs.
3). Assessed by comparing generated outputs with verified scientific data~\citep{guo2024can,bushuiev2024massspecgym}. 

\textbf{Multi-Dimensional Integration:} Use complementary benchmarks to capture performance holistically across different dimensions.
\\
\hline
\textbf{ Governance Infrastructure} & 
\textbf{Transparency: }In the chemistry domain, models must provide interpretable outputs, highlighting key chemical features influencing predictions and annotating data sources such as PubChem~\citep{pubchem2021} or ChEMBL~\citep{chembl2019}. Confidence scores and error margins should accompany predictions to ensure reliability.

\textbf{Governance:} It requires strict validation of outputs against experimental data, safeguards against misuse (e.g., toxic compound generation), and compliance with safety standards like EPA(Environmental Protection Agency)) and REACH(Registration, Evaluation, Authorization, and Restriction of Chemicals). Regular audits by chemists and toxicologists are essential to refine predictions and maintain ethical and safety standards.
 \\ 
\hline
\end{tabular}}
\caption{Application of the 3H framework to the development and application of chemistry foundation models.}
\label{app:3h_framework_chemistry}
\end{table*}

\clearpage

\end{document}